\documentclass[noacm,sigconf]{acmart}

\renewcommand\footnotetextcopyrightpermission[1]{} 
\AtBeginDocument{%
  \providecommand\BibTeX{{%
    \normalfont B\kern-0.5em{\scshape i\kern-0.25em b}\kern-0.8em\TeX}}}

\acmConference[]{}{}{}
\acmBooktitle{}
\acmPrice{1}
\acmDOI{}
\acmISBN{}

\begin{document}

\title{Hypothesis Testing Approach to Detecting Collusion in Competitive Environments}

\author{Pedro Hespanhol}
\affiliation{%
	\department{Industrial Engineering and Operations Research}
  \institution{University of California, Berkeley, CA}
}
\email{pedrohespanhol@berkeley.edu}

\author{Anil Aswani}
\affiliation{%
	\department{Industrial Engineering and Operations Research}
  \institution{University of California, Berkeley, CA}
}
\email{aaswani@berkeley.edu}


\begin{abstract}
There is growing concern about tacit collusion using algorithmic pricing, and regulators need tools to help detect the possibility of such collusion. This paper studies how to design a hypothesis testing framework in order to decide whether agents are behaving competitively or not. In our setting, agents are utility-maximizing and compete over prices of items. A regulator, with no knowledge of the agent's utility function, has access only to the agents' strategies (i.e., pricing decisions) and external shock values in order to decide if agents are behaving in competition according to some equilibrium problem. We leverage the formulation of such a problem as an inverse variational inequality and design a hypothesis test under a minimal set of assumptions. We demonstrate our method with computational experiments of the Bertrand competition game (with and without collusion) and show how our method performs. 
\end{abstract}




\maketitle

\section{Introduction}

Algorithmic pricing \cite{borenstein2004rapid,chen2016empirical} is increasingly used due to the growth of internet sales channels, but there is concern that these algorithms will lead to tacit collusion that harms consumers \cite{salcedo2015pricing,ezrachi2017artificial,calvano2018artificial,calvano2019algorithmic}. The current situation is unique in that, though it poses challenges for regulators because of the difficulty in detecting tacit collusion by algorithms, there is a large amount of real-time pricing and purchase data available for analysis by regulators. This paper is motivated by these pressing issues, and represents an initial step towards answering the question of how a regulator could detect algorithmic collusion from a large corpus of financial data.

The most closely related literature looks at collusion in auctions \cite{porter1993detection,baldwin1997bidder,pesendorfer2000study,bajari2003deciding}. Some work \cite{porter1993detection,pesendorfer2000study} conducted a statistical analysis of bidding data from situations where collusion is known to have occurred, and found that collusion leads to less aggressive bidding, higher prices for consumers, and increased correlation in bids. Other work \cite{porter1993detection,baldwin1997bidder,bajari2003deciding} uses econometrics to detect collusion, and is derived using analyses that assume (perfect) equilibrium behavior. However, assuming perfect equilibrium behavior is too stringent because it requires no model mismatch between the econometrics method and the bidders, and it requires bidders to be exactly accurate in the optimality of their bids. So these approaches generally lead to too many false positives when trying to detect collusion.

This paper proposes a hypothesis testing framework for detecting collusion between agents in competitive environments. The significant difference in our work is we allow the agents to not be in perfect equilibrium. Instead, we presuppose that agents typically choose actions close (but not exactly equal) to equilibrium but that they will also occasionally choose actions far from equilibrium. This weaker assumption partially mitigates model mismatch because it does not require any data without collusion to exactly match the equilibrium, and it also eliminates the need for agents to be exactly accurate in the optimality of their strategies. These ideas will be made more precise when we present our mathematical model.

\subsection{Estimation in Equilibrium Models}

Many competitive and cooperative environments can be analyzed using equilibrium models, where concepts like the Nash equilibrium are used to study agents' strategic behavior \cite{neck1987conflict,farrell1988communication}. Those models typically contain primitives -- such as agents' private information, utility functions, and strategy spaces -- that often are not known to an outside regulator or designer who then needs to estimate those elements in order to, for example, design a mechanism of interaction in order to induce a particular behavior or outcome. In this paper we use \emph{strategy space} and \emph{action space} interchangeably. 

However, the estimation of such primitives in equilibrium problems is quite challenging. In this work, we consider the case where the agents' actions are observed by the regulator and their strategy spaces are known. Hence the estimation problem lies solely on the agents' utility functions and personal information. One line of work is of structural estimation methods \cite{allon2011much,bajari2007estimating,rust1994structural}, which estimate parametric utility functions by observing agents acting in equilibrium. Those approaches assume a ``ground-truth'' form of the utility function, and derive necessary conditions based on constrained optimization in order to formally derive estimators of the parameters. Another related line of work are surrogate methods \cite{yang2007vcg,johari2009efficiency,farhadi2018surrogate} that, while not strictly estimation methods, elicit information from the agents themselves about sensitivities over strategies (i.e., about derivatives of their utility functions with respect to strategies) by providing the agents with a common surrogate function. Those methods can induce the appropriate equilibrium behavior even though the agents' utility functions are not known.

Lastly, estimation can be formulated as an inverse variational inequality \cite{harker1990finite,bertsimas2015data}. Such methods use the fact that equilibrium problems are a special case of variational inequalities, in order to pose an inverse optimization problem \cite{ahuja2001inverse} where the solutions (i.e., the equilibrium strategies) are provided via samples, and the goal is to estimate the problem's parameters that generate such solutions. This approach is powerful as it does not require a ``ground-truth'' model, and under some conditions can produce good approximate equilibrium behavior even when the parametric form of the utility functions is misspecified. However, these methods based on inverse optimization can encounter difficulties when the data gathered is noisy \cite{aswani2018inverse,aswani2019statistics}, which is what happens in most applications.

In this paper, we build on the framework of \cite{bertsimas2015data}. However, we will not assume that equilibrium actions are provided to the regulator via samples. Instead, we develop a method that given some arbitrary samples of actions is able to identify whether these samples came from agents acting in (approximate) equilibrium or not. Hence our work is more tied to coalition-detection in equilibrium games, in the field of economics, and to hypothesis testing in statistics. 

\subsection{Coalition Detection}
 
It is common to assume agents behave in equilibrium; however this is not observed in several applications \cite{bernheim1987coalition}. In fact, agents often collude, exchange information, and form sub-groups -- instead of cooperating as a whole \cite{staatz1983cooperative, greenberg1994coalition}. This poses a problem for estimation methods, which often assume the observed strategies come from an specific type of behavior (e.g., total cooperation or competition). The work \cite{saad2009coalitional, saad2011coalitional} characterizes coalitions in games and provides conditions where the establishment or not of coalitions can be tested. However the problem of coalition formation can also appear where agents play imperfectly and are learning the primitives of the environment while acting on it. Work done by \cite{li2004investigating} gives evidence that when agents employ learning algorithms in a competitive environment, such algorithms ``learn'' to implicitly collude, even if collusion is not part of the agents' plans.
 
 In this paper, we formulate the problem of identifying whether or not the observed actions come from agents in total competition or not. Instead of identifying the coalition itself based on structural properties of the game, we instead use a data-driven approach where a Kolmogorov-Smirnov hypothesis test \cite{massey1951kolmogorov,lilliefors1969kolmogorov} is used in order to accept or reject the hypothesis that the strategies observed by the regulator come from agents in equilibrium. This novel approach is powerful in the sense that is independent of the type of utility functions considered and makes very mild assumptions on the \emph{a priori} behavior of the agents, leveraging both the variational inequality formulation and the power of hypothesis testing.
 
\subsection{Outline}

In Section \ref{sec2}, we describe our (approximate) equilibrium model and develop the corresponding hypothesis testing framework. In Section \ref{sec3}, we present computational experiments based on a constrained Bertrand-type game to showcase the performance of our method when agents are acting in competition and in collusion. Lastly, we conclude in Section \ref{sec4} with a discussion on the potential future applications of our model and method. We also discuss open theoretical questions that are motivated by our work in this paper.

\section{Problem Setting}

\label{sec2}
We describe the agents' model, the estimation setting the regulator faces as it observes the actions taken by agents, and present our hypothesis testing framework for detecting collusion.

\subsection{Model Description}
Suppose two agents are engaged in a Bertrand-type game, competing over prices of certain items (e.g., products or airline tickets). Let $p_{i} \in \mathbb{R}^{q}$ be the price vectors of agent $i \in \{1,2\}$ over $\{1,...,q\}$ items, and note we use the terms \emph{prices} and \emph{strategies} interchangeably, as the strategy of each agent consists solely on the prices over the items. We assume the strategy space of agent $i$ is 
\begin{equation} \label{eq:cone}
\mathcal{P}_{i} = \{p \in \mathbb{R}^{q}: Ap =b, p \in K\}
\end{equation}
where $A$ is a $m\times q$ matrix, $b$ is a $m$-vector, and $K$ is a closed convex cone. Hence the strategy space is a cone given in standard form. In addition, each agent has their own utility function
\begin{equation}
\textstyle\textbf{U}_{i}(p_{1},p_{2}, \mu; \theta_{i}) = p_{i}D_{i}(p_{1},p_{2}, \mu; \theta_{i})
\end{equation}
where $D_{i}(p_{1},p_{2}, \mu; \theta_{i})$ is the agent's demand function, which depends on both price vectors and is parametrized by the vector $\theta_i \in \mathbb{R}^{d}$, which we assume to be the private information of each agent. Lastly, the demand function also depends on $\mu$, which is a common shock value disturbance that we assume to be a bounded disturbance with small magnitude (to be made precise in next section). The goal of this shock vector is to represent uncertainties that may affect the demand and are outside of the agents' control. 

The goal of each agent is to select a feasible price $p_{i} \in \mathcal{P}_{i}$ such that its utility function is maximized. We focus our analysis to the Nash equilibrium of the resulting game:

\begin{definition} A strategy profile $(p^{*}_{1},p^{*}_{2})$ is a Nash equilibrium if each agent plays the ``best-response'' to the other, namely if
\begin{equation} \label{eq:NE}
p^{*}_{i} \in \arg\max_{p_{i} \in \mathcal{P}_{i}}\textstyle\textbf{U}_{i}(p_{1},p^{*}_{2}, \mu; \theta_{i}), \mathrm{for }\ i \in \{1,2\}
\end{equation}
\end{definition}
The (pure-strategy) Nash Equilibrium may not be an adequate solution concept for this constrained environment of the Bertrand-type game, as it may not exist \cite{edgeworth1925pure}. However, we will focus on the case where each agent plays imperfectly. Namely, we assume that before playing the game, a gap value $\epsilon$ is sampled from a (known to the agents and regulator) parametric distribution $\mathcal{D(\phi)}$ with (unknown to the regulator but known to the agents) parameter $\phi$. Next, both agents pick a strategy vector $(p_1,p_2)$ that is an $\epsilon$-approximate Nash equilibrium of the Bertrand game. To formalize this notion of approximate equilibrium, we will use the characterization based on variational inequalities presented in \cite{bertsimas2015data}.
\begin{definition}
Given a function $\textbf{f}:\mathbb{R}^{q} \rightarrow :\mathbb{R}^{q}$ and a non-empty set $\mathcal{F} \in \mathbb{R}^{q}$, the problem of finding the point $p^{*}$ such that
\begin{equation} \label{eq:VI}
\textbf{f}(p^{*})^{\top}(p-p^{*}) \geq 0, \mathrm{for }\ p \in \mathcal{F}
\end{equation}
is called the variational inequality problem $VI(\textbf{f},\mathcal{F})$. 
\end{definition}
It turns out that several problems can be formulated as variational inequalities (We refer to \cite{harker1990finite} for an in-depth characterization). In particular, if we let $\mathcal{F} = \mathcal{P}_{1} \times \mathcal{P}_{2}$ and we let
\begin{equation} \label{eq:f_def}
\textbf{f}(p) = \begin{bmatrix} \textbf{f}_{1}(p_1,p_2) \\ \textbf{f}_{2}(p_1,p_2)\end{bmatrix}= \begin{bmatrix} -\nabla_{1}U_{1}(p_1,p_2,\mu;\theta_1) \\ -\nabla_{2}U_{2}(p_1,p_2,\mu;\theta_2) 
\end{bmatrix}
\end{equation}
where $\nabla_{i}$ is the gradient w.r.t. $p_i$, then solving $VI(\textbf{f},\mathcal{F})$ from (\ref{eq:VI}) is equivalent to finding the Nash equilibrium (\ref{eq:NE}). We thus have the following definition for approximate Nash equilibrium:
\begin{definition}
A strategy profile $(\bar{p}_1,\bar{p}_2)$ is an $\epsilon$-approximate Nash equilibrium if and only if
\begin{equation} \label{eq:eps_VI}
\textbf{f}(\bar{p})^{\top}(p-\bar{p}) \geq -\epsilon, \mathrm{for}\ p \in \mathcal{F}.
\end{equation}
\end{definition}

It will suit our purposes to formulate the above approximate variational inequality problem as a (convex) optimization problem. This can be done under technical regularity conditions that ensure constraint qualification holds (e.g., Slater's condition). 
\begin{theorem}\cite{bertsimas2015data}
 Let $\mathcal{F} = \mathcal{P}_{1} \times \mathcal{P}_{2}$, where $\mathcal{P}_{i}$ is given by (\ref{eq:cone}), for $i \in \{1,2\}$. Let $\textbf{f}$ be given by (\ref{eq:f_def}). If $\mathcal{F}$ satisfies constraint qualification (e.g., Slater's condition), then a strategy profile $(\bar{p}_1,\bar{p}_2)$ is an $\epsilon$-approximate Nash equilibrium if and only if
\begin{equation} \label{eq:VI_feas}
\exists y_{1},y_{2} \in \mathbb{R}^{m}: \begin{cases}
A^{\top}_{i}y_{i} \leq_{C} \textbf{f}_{i}(\bar{p}_{1}, \bar{p}_{2}) , \mathrm{for}\ i \in \{1,2\} \\
\sum_{i=1}^{2}\textbf{f}_{i}(\bar{p}_{1}, \bar{p}_{2})^{\top}\bar{p}_{i} - b^{\top}_{i}y_{i} \leq \epsilon
\end{cases}
\end{equation}
where we use the symbol ``$\leq_{C}$'' to denote conic inequalities.
\end{theorem}

Next we assume that given some $\epsilon \sim \mathcal{D(\phi)}$, the agents solve the above feasibility problem in order to select the prices. In particular, we assume that the agents solve the above problem where the second inequality is replaced by an equality constraint -- that is the selected strategies satisfy condition (\ref{eq:eps_VI}) with equality. We will not focus on how such prices are achieved, that is, how the agents learn to play the $\epsilon$-approximate Nash Equilibrium strategies (we refer to \cite{li2004investigating} for a discussion about learning in cooperative games). Instead, we will focus on the following estimation problem faced by an external regulator: Given a sequence of observed prices and shocks $\{(p^{j}_{1},p^{j}_{2}, \mu^{j})\}_{j=1}^{N}$, the regulator would like to ascertain whether or not agents are playing according to $\epsilon$-approximate Nash Equilibrium or not. In this setup, the private information vectors $(\theta_{1},\theta_{2})$ of each agent are so-called \emph{nuisance parameters} for the regulator (i.e., they require estimation even though they are not of primary interest). To that end, the regulator will construct estimates $(\hat{\theta}_{1}, \hat{\theta}_2)$ of the private information vectors and residual estimates $\hat{\epsilon}_{j}$ for each observation tuple $j \in \{1,...,N\}$ by solving the inverse variational problem given by
\begin{align} \label{eq:inv_opt}
\min_{\hat{\theta},y, \hat{\epsilon}}\ & L(\hat{\epsilon}^{1},...,\hat{\epsilon}^{N}) \\
\mathrm{s.t.}\ & A^{\top}_{i}y^{j}_{i} \leq_{C} \textbf{f}_{i}(p^{j}_{1}, p^{j}_{2}), \mathrm{for}\ i \in \{1,2\}, j \in \{1,...,N\} \\
& \textstyle\sum_{i=1}^{2}\textbf{f}_{i}(p^{j}_{1}, p^{j}_{2})^{\top}p^{j}_{i} - b^{\top}_{i}y^{j}_{i} = \hat{\epsilon}^{j}, \mathrm{for}\ j \in \{1,...,N\} 
\end{align}
where $L(\hat{\epsilon}^{1},...,\hat{\epsilon}^{N})$ is some loss function over the residual estimates. We assumed that the regulator knows the distribution $\mathcal{D}(\phi)$, but does not know $\phi$. Hence the loss function can be written, for example, as the negative log-likelihood as a function of $\phi$ \cite{severini2000likelihood}. We note in this optimization problem, the prices are given by our $N$ samples, and we seek to select a $\hat{\theta}$ such that the resulting utilities form an approximate Nash equilibrium for every sample collected, where the computed $\hat{\epsilon}_{j}$ are our residual estimates of $\epsilon$.

Lastly, in order to make a decision as to whether or not the observed prices are in approximate Nash equilibrium, the regulator will formulate a hypothesis test over the computed residuals.

\subsection{Hypothesis Testing Framework}

In order to formalize the hypothesis testing framework, we begin by describing the temporal sequence of events under consideration:
\begin{enumerate}
\item Both agents and regulator observe $\mu$, the shock variable.

\item The agents solve the feasibility problem (\ref{eq:VI_feas}) for some $\epsilon \sim\mathcal{D}(\phi)$. The strategies $(p_{1},p_{2})$ are selected to exactly be an $\epsilon$-approximate Nash Equilibrium.

\item The regulator observes the strategies $(p_{1},p_{2})$ and records it.

\item Steps 1-3 are repeated $N$ times and the regulator collects the sample tuples $\{(p^{j}_{1},p^{j}_{2},\mu^{j})\}_{j=1}^{N}$.

\item The regulator solves the inverse variational problem (\ref{eq:inv_opt}) for some parametric utility functions and computes the estimated residuals $\hat{\epsilon}^{1},...,\hat{\epsilon}^{N}$.

\item The regulator uses those residuals to perform a Kolmogorov-Smirnov test (to be defined next).
\end{enumerate}

We note that the regulator does not know the true utility functions of the agents. Importantly, our approach is partially amenable to parametric form misspecification because non-colluding agents are not required to be in perfect equilibrium. In other words, some amount of the $\epsilon^j$ are meant to capture model misspecification.

Step 6 is conducted as follows: The regulator will use the computed estimated residuals to perform a hypothesis test to determine if the $\hat{\epsilon}^{1},...,\hat{\epsilon}^{N}$ come from the distribution $\mathcal{D}(\phi)$. However, even though the regulator knows the distribution's parametric form, they do not know the underlying parameter $\phi$. Hence, hypothesis tests such as the standard Kolmogorov-Smirnov test are not applicable since they require knowing the true underlying parameters of the distribution under the null hypothesis. Therefore, we we will resort to the Lilliefors variation of the Kolmogorov-Smirnov test \cite{lilliefors1969kolmogorov}. We first compute the empirical cumulative distribution function $\textstyle\hat{F}_{N}(d) = \frac{1}{N}\sum_{j=1}^{N} \mathbb{I}(\hat{\epsilon}_{j} \leq d)$, where $ \mathbb{I}(\cdot)$ is an indicator function. Then the regulator computes some estimate $\hat{\phi} = g(\hat{\epsilon}^{1},...,\hat{\epsilon}^{N})$ and computes the cumulative distribution function $\bar{F}_{N}(d) = F_{\mathcal{D}(\hat{\phi})}(d)$, where $F_{\mathcal{D}(\hat{\phi})}(d)$ is the cumulative distribution function of a random variable of distribution $\mathcal{D}(\hat{\phi})$. Lastly, the regulator computes the test statistic $D^{*} = \max_{d} |\hat{F}_{N}(d) - \bar{F}_{N}(d)|$. The null $H_0$ and alternative $H_1$ hypotheses for our test are
\begin{equation}
\mathcal{H}: \begin{cases}
H_{0}: &\text{The agents are behaving in an $\epsilon$-approximate}\\
 &\text{equilibrium where $\epsilon \sim\mathcal{D}(\phi)$} \\
H_{1}: &\text{Otherwise}
\end{cases}
\end{equation}
And the decision of whether to accept or reject the null hypothesis is made using the decision-rule
\begin{equation} \label{eq:HT}
\begin{cases}
\text{reject } H_{0}: &\text{if }D^{*} \geq \tau(N) \\
\text{accept } H_{0}: &\text{if }D^{*} < \tau(N) \\
\end{cases}
\end{equation}
where $\tau(N)$ is some threshold from the Lilliefors variation of the Kolmogorov-Smirnov test \cite{lilliefors1969kolmogorov} and which is based on the number of samples collected and the desired significance level $\alpha$.

\section{Computational Experiments}
\label{sec3}

Here, we analyze the performance of our approach in a Bertrand competition environment. We first detail the experiment setting and then proceed to the numerical experiments and analysis.

\subsection{Experiment Setting}

We showcase our method in a setting where two agents compete over a single item and need to set their respective prices in the Bertrand-game environment. Each agent's true demand function has the following form:
\begin{equation}
\textstyle\bar{D}_{i}(p_{1},p_{2}, \mu; \bar{\theta}_{i}) = \bar{\theta}_{i,0} + \sum_{j=1}^{2}p_{j}\bar{\theta}_{i,j} + \bar{\theta}_{i,3}\mu + \eta_{i}
\end{equation}
where $\bar{\theta}_{i}$ is the agent's private information vector, and we use the term $\eta_{i}$ to encompass unmodeled terms of the dynamics. Furthermore, we assume the set of feasible price vectors belong to the polyhedral set
\begin{equation}
\label{eq:abs3}
\begin{aligned}
\mathcal{P} = \{(p_1,p_2) \in \mathbb{R}^{2}: \text{ }
&0 \leq p_{1} \leq \bar{p}, 0 \leq p_{2} \leq \bar{p} \}
\end{aligned}
\end{equation}
where 
$\bar{p}$ is an upper-bound on each price. We consider the case where $\epsilon$ is drawn from an exponential distribution $\epsilon \sim \exp(\bar{\lambda})$. 

We assume that the regulator observes the shock $\mu$ but does not observe $\eta_{i}$. Hence, the regulator forms the following demand estimate given some estimate $\hat{\theta}$:
\begin{equation}
\textstyle\textbf{D}_{i}(p_{1},p_{2}, \mu; \hat{\theta}) =  \hat{\theta}_{i,0} + \sum_{j=1}^{2}p_{j}\hat{\theta}_{i,j} + \hat{\theta}_{i,3}\mu
\end{equation}
Following the steps described in the previous section, the regulator collects the sample tuples $\{(p^{j}_{1},p^{j}_{2},\mu^{j})\}_{j=1}^{N}$ and forms the optimization problem (\ref{eq:inv_opt}) with the loss function $L(\hat{\epsilon}^{1},...,\hat{\epsilon}^{N})$ being the negative log-likelihood of the underlying exponential distribution. Note that in the negative log-likelihood the $\lambda$ term is decoupled from the other terms because of the particular mathematical form of the density of an exponential distribution. As a result, we do not need to include $\lambda$ in the inverse variational problem.

In order to make the presentation of the final optimization problem clear, we define the marginal utility function for each agent (as considered by the regulator) to be
\begin{multline}
m_{i}(p_1,p_2,\mu;  \hat{\theta}_i) = p_{i}\frac{\partial}{\partial p_i}\textbf{D}_{i}(p_{1},p_{2}, \mu; \theta_{i}) + \textbf{D}_{i}(p_{1},p_{2}, \mu; \theta_{i}) = \\
\textstyle p_i\hat{\theta}_{i,i} + \hat{\theta}_{i,0} + \sum_{j=1}^{2}p_{j}\hat{\theta}_{i,j} + \hat{\theta}_{i,3}\mu.
\end{multline}
In addition, we impose some structure to the fitted utility functions: (1) we normalize the fitted utility functions; (2) we enforce that the marginal utilities of each agent decrease as they increase their own prices (on the observed data); and (3) we enforce an additional constraint that sets the dual variable $y^{j}_{i}$ to zero if the observed price $p^{j}_{i}$ is strictly less than the upper bound $\bar{p}$. Recalling the definition of $\textbf{f}(p)$ in (\ref{eq:VI}), the optimization problem becomes
\begin{align}
\min_{\hat{\epsilon},y,\theta_1,\theta_2}\ &\textstyle\sum_{j=1}^{N} \hat{\epsilon_{j}}\\
 \mathrm{s.t. }\ & y^{j}_{i} \geq m_{i}(p^{j}_1,p^{j}_2,\mu^{j},\theta_1), \mathrm{for}\ i \in \{1,2\}, j \in \{1,...,N\} \\
 &\textstyle\bar{p}\sum_{i=1}^{2}(y^{j}_{i}) - \sum_{i=1}^{2}p^{j}_{i}m_{i}(p^{j}_1,p^{j}_2,\mu^{j},\theta_i)  = \hat{\epsilon}_{j},\\
 & \quad \quad \quad \quad \quad \quad \quad \quad \quad \quad \quad \quad \mathrm{for}\  j \in \{1,...,N\} \nonumber\\
 & m_{i}(1,1,0,\theta_i) = m_{i}(1,1,0,\bar{\theta}_i), \mathrm{for}\  i \in \{1,2\} \label{eq:norm}\\
 & y^{j}_{i} = 0, \mathrm{for}\  i \in \{1,2\}, j \in \{1,...,N\}\ \mathrm{s.t. }\ p^{j}_{i} < \bar{p} \\
 & \theta_{i,i} \leq 0, \mathrm{for}\  i \in \{1,2\} \label{eq:noninc} \\
 &\hat{\epsilon}^{j} \geq 0, \mathrm{for}\  j \in \{1,...,N\} \label{eq:dual}\\
 & y^{j}=( y^{j}_{1}, y^{j}_{2}) \geq 0, \mathrm{for}\  \forall j \in \{1,...,N\} 
\end{align}
where (\ref{eq:norm}) are the normalization constraints, in which the marginal utility of both agents when there is no external shock and the prices are set to unity is equal to the true marginal at that point.  (Note we could have set these normalization constraints to any other suitable positive value without affecting the results.  Different normalization may yield different models that can be used to explain the same observed data. This phenomenon is common in inverse optimization problems, as discussed in detail in \cite{bertsimas2015data,aswani2018inverse}.) Equation (\ref{eq:noninc}) ensures that the fitted marginal functions decrease as the agents increase their own prices. (This constraint is obtained after some arithmetic by requiring that $m_{1}(p_1,\cdot,\cdot; \theta_i)$ and $m_{2}(\cdot,p_2,\cdot; \theta_i)$ decrease as $p_1$ and $p_2$ increase, respectively, on the observed data.) The ``dual'' vector $y$ is associated with the constraints (\ref{eq:abs3}) and (\ref{eq:dual}) ensures that if the the observed prices are not on the boundary of the feasible region $\mathcal{P}$ then the associated dual variable is set to zero. We note that (\ref{eq:dual}) has a very subtle implication in the optimization problem above: The very natural notion that dual variables are zero once their associated constraint is non-binding is not enforced at all by the original formulation in (\ref{eq:inv_opt}). If the prices are sampled in perfect Nash equilibrium (that is $\epsilon = 0$), then as argued in \cite{bertsimas2015data} the formulation in (\ref{eq:inv_opt}) is able to recover exactly the true parameters $\theta_i$ and the computed residuals are exactly zero. However, in our scenario prices are obtained in approximate equilibrium (i.e., $\epsilon >0$). Hence if complimentary slackness (\ref{eq:dual}) is not enforced explicitly then the computed residuals will present bias -- namely they will be ``shrunk'' since the formulation (\ref{eq:inv_opt}) could achieve smaller values for the residuals by setting the dual variables to be positive, even though the sampled prices are in the interior of the feasible region. 

\begin{table}
\caption{\label{table_1}Numerical Results for Scenario 1 (Competing)}
\begin{center}
\begin{tabular}{c  c  c  c  c} 
\toprule
N & $D^{*}$ & $\tau(N)$ & $\hat{\lambda}$ & Decision \\
\midrule
10 &  0.317 & 0.325  & 33.7 & Competing  \\
20 & 0.206 & 0.234 & 27.93 & Competing   \\
30 & 0.120 & 0.192 & 20.83 & Competing\\ 
40 & 0.069 & 0.168 & 21.43 & Competing\\
50 & 0.089 & 0.150 & 19.99 & Competing\\
100 & 0.070 & 0.106 & 18.80 & Competing\\
200 & 0.031 & 0.075 & 18.62 & Competing\\
500 & 0.022 & 0.047 & 20.01 & Competing\\
\bottomrule
\end{tabular}
\end{center}

\end{table}

Recall that the objective function follows from the negative log-likelihood of exponential distribution, where we dropped the term $N/\lambda$ since it does not impact the optimization. After solving the problem above, we compute the MLE estimate $\hat{\lambda}_{MLE} = \textstyle(\frac{1}{N}\sum_{j=1}^{N} \hat{\epsilon}_{j})^{-1}$ of $\lambda$. Then we let $\bar{F}_{N}(d) = F_{\exp(\hat{\lambda}_{MLE})}(d)$ and conduct the Lilliefors hypothesis test (\ref{eq:HT}). To illustrate the performance of the hypothesis testing we will simulate the process under two scenarios:\\

\newlength{\myl}
\settowidth{\myl}{\quad}

\begin{center}
\begin{minipage}{0.9\columnwidth}
\textbf{Scenario 1:} Agents are competing over prices, i.e.: they solve the feasibility problem (\ref{eq:VI_feas}) after observing the shock variable $\mu$ and the value of $\epsilon$.\\
\end{minipage}

\begin{minipage}{0.9\columnwidth}
\textbf{Scenario 2:} Agents are colluding, i.e.: instead of solving the feasibility problem (\ref{eq:VI_feas}), they maximize the sum of both utility functions up to a $\epsilon$ optimality gap.\\
\end{minipage}
\end{center}

\noindent Hence for Scenario 2, prices are generated by solving 
\begin{equation}
(p^{j}_1,p^{j}_2) = \arg\max_{(p_{1},p_{2}) \in \mathcal{P}} \textstyle\sum_{i=1}^{2} p_{i}D_{i}(p_1,p_2,\theta_i,\mu_{j})
\end{equation}
for $j \in \{1,...,N\}$. In the next subsection, we present numerical simulations of these two scenarios and show how the regulator rejects/does not reject the null hypothesis as the agents change their behavior from competition to collusion.

\subsection{Computational results}

We let $\bar{\lambda} = 20$ and chose $\bar{\theta}_{1} = [10,-1,0.5,1]$ and $\bar{\theta}_{2} = [8,0.4,-3.0,1]$ to be the agents' true private information vectors. The shock values were generated by $\mathcal{N}(5, 1)$, and we fix the upper-bound $\bar{p} = 8.0$ on the prices. Furthermore, we fix our significance level $\alpha = 0.05$. The threshold $\tau(N)$ for the hypothesis testing is obtained by the table presented in \cite{lilliefors1969kolmogorov}. Lastly, we let $\eta_j$ for $j=\{1,2\}$ be sampled from $\mathcal{N}(0,1)$. For the first scenario, the approximate equilibrium prices need to be generated by solving (\ref{eq:VI_feas}). In our test case, we generate approximate equilibrium prices via the algorithm described in the appendix. The results for Scenario 1 are summarized in Table \ref{table_1}.

\begin{figure}
\includegraphics[clip,scale=0.5]{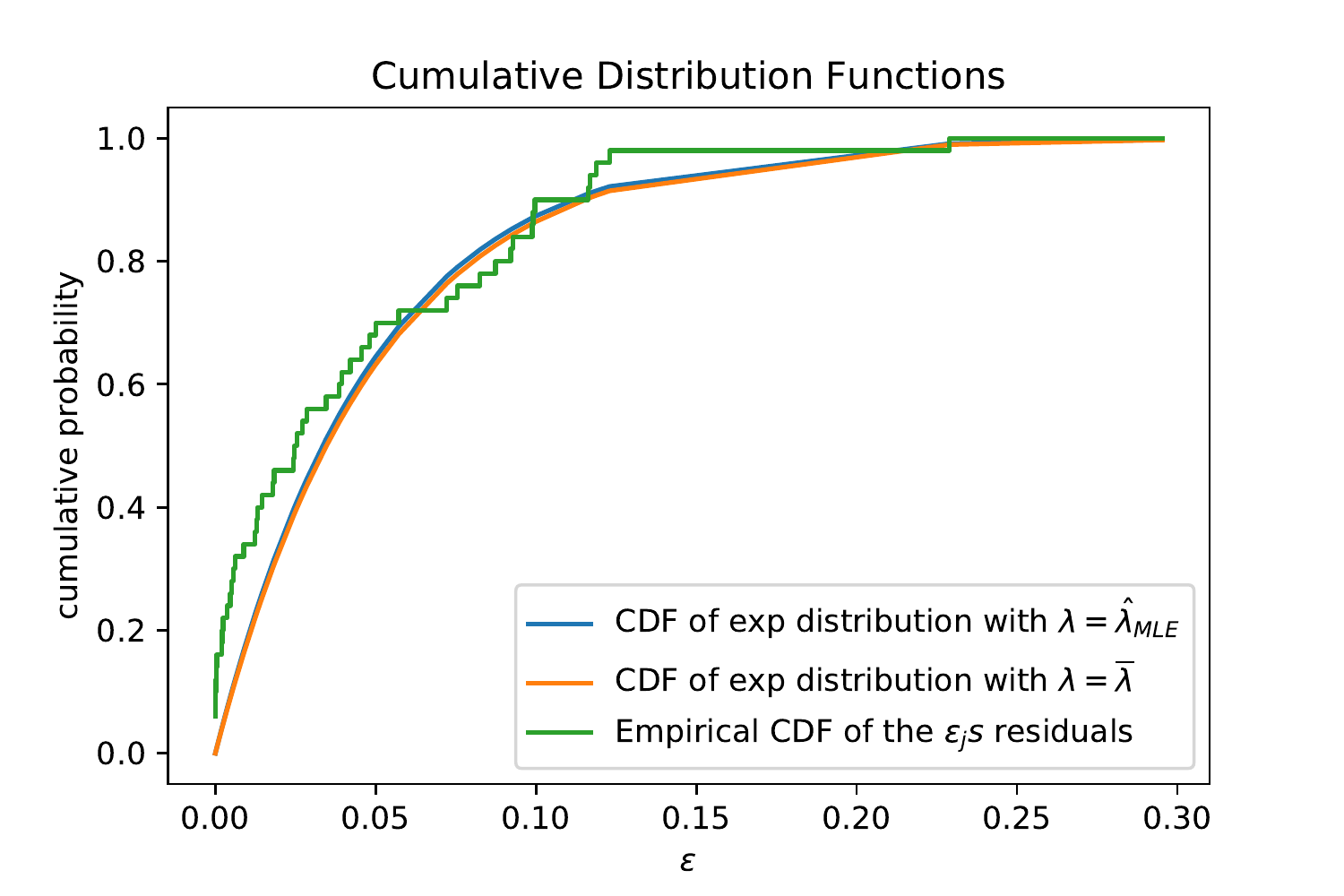}
\caption{\label{fig:one}Comparing CDF's of Residuals For Scenario 1}
\Description{Comparing CDF's of Residuals For Scenario 1}
\end{figure}

When agents are competing (i.e., acting under the specifications of the null hypothesis), a false positive (i.e., decision of collusion occurring) was not seen in the experiments. This is not surprising because we set $\alpha = 0.05$ and each row in the table corresponds to a single numerical experiment. If we repeated these experiments, we would expect to see a close to $\alpha$ fraction of them report a false positive. Also, note we are able to recover the correct estimate of $\lambda$ for the underlying distribution generating the residuals. This is highlighted in Figure \ref{fig:one}, where we plot the $\epsilon_{j}'s$ samples from $\exp(20)$ and the computed residual estimates $\hat{\epsilon}_{j}'s$ by the regulator after solving the optimization problem, for sample size equal to $50$.

\begin{figure}
\includegraphics[clip,scale=0.5]{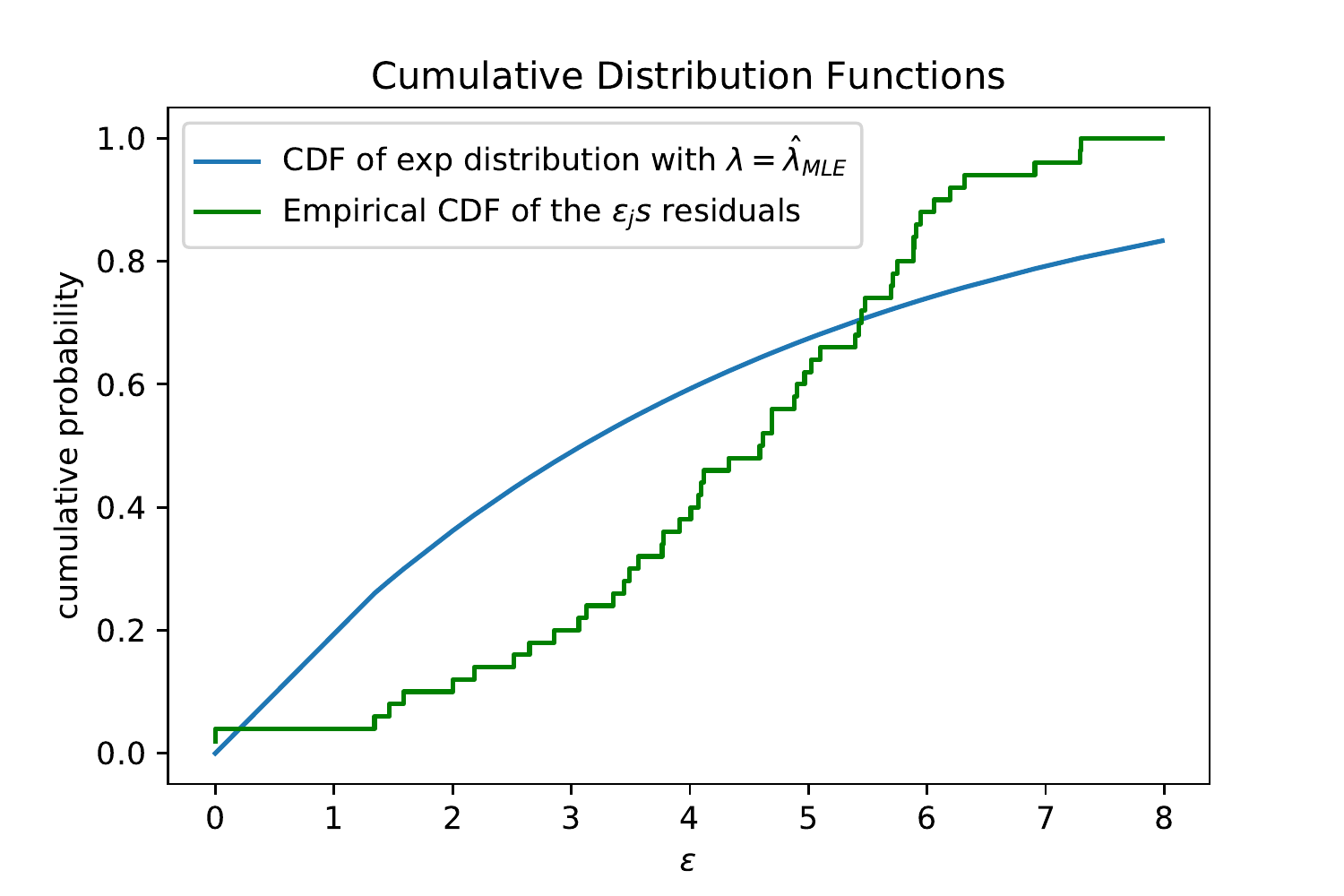}
\caption{\label{fig:two}Comparing CDF's of Residuals For Scenario 2}
\Description{Comparing CDF's of Residuals For Scenario 2}
\end{figure}

For the second scenario we generate prices by solving an aggregate problem where we sum both agents' utilities to compute the prices. In Table \ref{table_2}, we see that the null hypothesis is rejected (i.e., decision of collusion occurring) for moderate and large sample sizes. The MLE estimate of $\bar{\lambda}$ is inaccurate as well since the agents are not behaving in approximate equilibrium. In Figure \ref{fig:two}, for $N=50$ we plot the empirical CDF of residual estimates $\hat{\epsilon}_{j}'s$ by the regulator in this scenario. Observe that when agents are cooperating instead of competing, the computed residuals are vastly different then their true values (we omit plotting the true cdf of $\exp(20)$ since the computed residuals are very large for this scenario). The null hypothesis that the agents are competing in equilibrium is rejected for almost all sample sizes, indicating that our method is able to identify when agents are not behaving in competition. We stress that rejecting the null hypothesis is not proof that agents are colluding, but rather gives some statistical evidence that suggests collusion is occurring. 


\begin{table}
\caption{ \label{table_2}Numerical Results for Scenario 2 (Colluding)}
\begin{center}
\begin{tabular}{c  c  c  c  c} 
\toprule
N & $D^{*}$ & $\tau(N)$ & $\hat{\lambda}$ & Decision \\
\midrule
10 &  0.261 & 0.325  & 0.12 & Competing  \\
20 & 0.263 & 0.234 & 0.11 & Colluding   \\
30 & 0.222 & 0.192 & 0.22 & Colluding\\ 
40 & 0.300 & 0.168 & 0.26 & Colluding\\
50 & 0.301 & 0.150 & 0.22 & Colluding\\
100 & 0.322 & 0.106 & 0.28 & Colluding\\
200 & 0.301 & 0.075 & 0.30 & Colluding\\
500 & 0.335 & 0.047 & 0.30 & Colluding \\
\bottomrule
\end{tabular}
\end{center}
\end{table}

\section{Conclusion and Future Work}

\label{sec4}

We proposed a hypothesis testing framework to decide whether agents are behaving competitively or not. In our setting, a regulator formulates an inverse variational problem in order to estimate the unknown private information vectors as well as estimate the residuals of the approximate equilibrium that arises from the agents' competition. Our setting is flexible as the regulator only requires access to prices and shock values. A future direction of work is to study the theoretical properties of our framework. We demonstrated our method in a simple two-player game with a polyhedral feasible action space. We stress our setting is more general and allows for any number of players with arbitrarily conic-representable sets, as long as they satisfy some regularity conditions. Another direction of future research is to apply our testing framework to the setting in \cite{calvano2018artificial,li2004investigating}, where groups of agents ``learn'' to collude instead of competing. This problem is more challenging but can be explored in the light of inverse variational problems and our framework.

\begin{acks}
This material is based upon work supported by the National Science Foundation under Grant CMMI-1847666, and by the UC Berkeley Center for Long-Term Cybersecurity.
\end{acks}

\bibliographystyle{ACM-Reference-Format}
\bibliography{Inverse}

\appendix

\section{Algorithm to Generate Approximate Equilibrium Prices}

A key part of our numerical simulations is to generate prices that are $\epsilon$-approximate equilibrium. In the general case, we need to solve the variational inequality formulation in (\ref{eq:VI_feas}). That problem is hard to solve in general, but tailored algorithms do exist \cite{giannessi2006equilibrium}. However, for our setting the feasible region $\mathcal{P}$ contains only bounds on the prices. Hence the problem becomes to find prices $(p_1,p_2)$ such that
\begin{equation} \label{eq:VI_feas_simple}
\exists y_{1},y_{2} \geq 0:\begin{cases}
y_{i} \geq D_{i}(p_1,p_2.\mu,\bar{\theta}_{i}) + p_i\theta_{i,i}, \mathrm{for}\ i \in \{1,2\} \\
\sum_{i=1}^{2}\bar{p}y_{i} -  p_i(D_{i}(p_1,p_2.\mu,\bar{\theta}_{i}) + p_i\theta_{i,i}) = \epsilon
\end{cases}
\end{equation}
Hence we can generate samples of $(p_1,p_2)$ by acceptation/rejection of samples based on the shock values $\mu$ and nuisance parameters $(\eta_1,\eta_2)$. First, we sample $\mu$ and $\eta_1,\eta_2$ according to their specified distributions. Then we solve the following system of nonlinear equations (via, for example, Newton's Method): $p_{i}D_{i}(p_1,p_2,\mu,\bar{\theta}_{i}) + (p_i)^{2}\theta_{i,i} = \frac{-\epsilon}{2}$, for $i \in \{1,2\}$. After solving this system, if $(p_1,p_2) \in \mathcal{P}$ then it means they are $\epsilon$-approximate solution to the variational inequality problem (since we can set both $y_1$ and $y_2$ to zero), and we accept the sample $(p_1,p_2,\mu)$. If $p_1<0$ or $p_2<0$, then we reject the sample. Now without loss of generality, suppose that $p_1 > \bar{p}$. Then we can set $p_1 = \bar{p}$ and let $y_1 = D_{1}(\bar{p},p_2.\mu,\bar{\theta}_{1}) + \bar{p}\theta_{1,1}$. Then by letting $y_2 = 0$ we solve for $p_2$ by $p_{2}D_{2}(\bar{p},p_2,\mu,\bar{\theta}_{i}) + (p_2)^{2}\theta_{2,2} = -\epsilon$. Lastly if $p_2 \geq 0$ and $y_1 \leq 0$, then we accept the sample $(p_1,p_2,\mu)$. In all other cases, we reject the sample. With this simple method, we can generate sample prices that are $\epsilon$-approximate equilibrium. By repeating the above $N$ times for each sampled $\epsilon_j$, we can generate all the samples necessary for the numerical simulation.

\end{document}